# New examples of metalloaromatic Al-clusters: $(Al_4M_4)Fe(CO)_3$ (M=Li, Na and K) and $(Al_4M_4)_2Ni$: Rationalization for possible synthesis

Ayan Datta and Swapan K. Pati [*]

*Theoretical Sciences Unit and Chemistry and Physics of Materials Unit, Jawaharlal Nehru Center for Advanced Scientific Research, Jakkur P. O, Bangalore-560064, India.*

**Abstract**

*Ab-initio* calculations reveal that all-metal antiaromatic molecules like $Al_4M_4$ (M=Li, Na and K) can be stabilized in half-sandwich complex: $(Al_4M_4)Fe(CO)_3$ and full-sandwich complexes of the type: $(Al_4M_4)_2Ni$. The formation of the full-sandwich complex $[(Al_4M_4)_2Ni]$ from its organometallic precursor depends on the stability of the organic-inorganic hybrid $(C_4H_4) Ni (Al_4Li_4)$.

The concept of aromaticity and antiaromaticity is of fundamental importance to the structural chemistry.[1-3] This concept has been recently extended from the organic molecules to metallic systems like Al and Sn.[4,5] Molecules like $Li_4Al_4$, $Na_4Al_4$ and their anions ($Li_3Al_4^-$ and $Na_3Al_4^-$) have been recently shown as the examples of first antiaromatic metal clusters due to their close resemblence with $C_4H_4$, from both structural and electronic point of view.[6] Also, very recently, we have shown that these clusters can be very well stabilized by complexation with 3d-metal atoms like Fe and Ni which results in metalloaromaticity in these $Al_4$-clusters with increase in the number of π-electrons from 4π to 6π.[7] This is very similar in concept to the onset of aromaticity in $C_4H_4$ on complexation, originally proposed by H. C. Longuet Higgins et al. and synthesized soon after by Criegee et al.[8] The counterions in $Al_4M_4$ (M=Li, Na and K) being highly electropositive, lose the electrons to the $Al_4$, thereby making the ring formally (-4) and thus a 4π antiaromatic system. $Al_4M_4$ can thus be considered as isolobal with $C_4H_4$ and is expected to follow similar reactivity as shown for $C_4H_4$.[9] The role of the counterions have however, not been investigated till now and in this work we critically examine their role and propose the methodologies for possible experimental synthesis through stepwise formation of half-sandwich all-metal complexes, hybrid organo-inorganic complexes and finally full-sandwich all-metal complexes. We also study the bonding aspects in the first mixed sandwich (hybrid) complexes of the type: $(Al_4M_4)(C_4H_4)Ni$. Interestingly, we find that these hybrid sandwich complexes have admixtured interactions with the transition metal atoms and organic/inorganic ligands and can be the candidates for future synthesis.

The structures considered for the present work: $(Al_4M_4)$-$Fe(CO)_3$, sandwich complexes of the type: $(Al_4M_4)_2$-Ni and the hybrid-sandwich complexes: $(Al_4M_4)(C_4H_4)$-Ni with M=Li, Na and K, were optimized based on the B3LYP/6-31G(d,p)+ method (see supporting information file for details).[10] The calculations for the fragmentation energy analysis and the HOMO-LUMO gaps were performed at the B3LYP/6-311G(d,p)++ level. The same level of calculations were also performed for the uncoordinated Al4M4 and C4H4. We first discuss the structural features in the uncoordinated ligands, Al4M4. There are some remarkable similarities in their structures with variation in the alkali metal ions. The ground state minimum energy form for all the ligands possess a C2h symmetry (See Fig. 1).

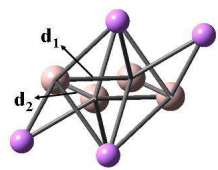

**Figure 1.** Ground state geometry for $Al_4M_4$. Bond-distances $d_1$ and $d_2$ (in Å) are 2.68, 2.56 for $Al_4Li_4$, 2.70, 2.59 for $Al_4Na_4$ and 2.67, 2.58 for $Al_4K_4$ respectively. See Supp. information file for coordinates and energies of these geometries.

Another very important feature to be noticed for these molecules is that the ground state geometry for all the ligands have a substantial bond length alteration (BLA) [0.124 Å for $Al_4Li_4$, 0.11 Å for $Al_4Na_4$ and 0.10 Å for $Al_4K_4$]. BLA is defined as the bond length difference between two subsequent Al-Al bonds in the central four-membered $Al_4$ ring. For their organic analogue, $C_4H_4$, the BLA is 0.245 Å. While, the magnitude of BLA in these clusters are smaller than that for the π-conjugated antiaromatic molecule, $C_4H_4$, it is more than that in the aromatic molecules like benzene (BLA=0) and thus these clusters are both ?-aromatic as well as ?-antiaromatic.[6a,11] We have calculated the nucleus-independent chemical shift (NICS) at the GIAO-B3LYP/ 6-311G(d,p)++ level for these molecules.[11] The NICS values for $Al_4Li_4$, $Al_4Na_4$ and $Al_4K_4$ are –11.55 ppm, -7.91 ppm and –7.72 ppm respectively. Interestingly, with variation in the size of the alkali metal, we find that the extent of NICS aromaticity decreases with the increase in the size of the counterion. This is understood from the fact that the ionization potentials in M decreases in the order, Li > Na> K. Thus, the extent of charge transfer is maximum in $Al_4K_4$ from the K ion to the $Al_4$ ring and is strongly ?-antiaromatic which reduces the magnitude of the ?-aromaticity. A similar conclusion is also derived by the magnitudes in the HOMO-LUMO gaps. The gaps for $Al_4Li_4$, $Al_4Na_4$ and $Al_4K_4$ are 1.45 eV, 1.25 eV and 0.74 eV respectively. The much smaller gap for $Al_4K_4$ signifies high reactivity, smaller chemical hardness and poor overall aromatic character.[13]

In the complex $(Al_4M_4)$-$Fe(CO)_3$, in all the cases with variation in M, the ligand binds strongly with the Fe atom forming an $\eta^4$ coordination. A very remarkable feature in these complexes is that in the $Al_4^{4-}$ rings, the BLA is very small [0.028 Å, 0.0345 Å and 0.041 Å in $(Al_4Li_4)Fe(CO)_3$, $(Al_4Na_4)Fe(CO)_3$ and $(Al_4K_4)Fe(CO)_3$ respectively].

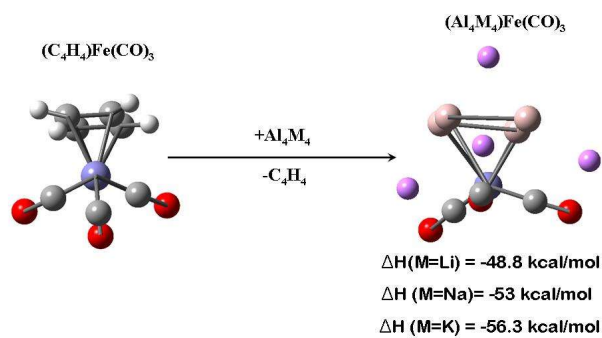

**Figure 2.** Substitution reactions in $(C_4H_4)Fe(CO)_3$ by $Al_4Li_4$, $Al_4Na_4$ and $Al_4K_4$ to produce $(Al_4Li_4)Fe(CO)_3$, $(Al_4Na_4)Fe(CO)_3$ and $(Al_4K_4)Fe(CO)_3$. Note that all these substitutions are highly exothermic. Small filled circles indicate the alkali metal ions.

This suggests that the ligand interacts with the d-orbitals of the Fe atom, with a transfer of 2 electrons from the metal to the ligand (MLCT) making the $Al_4$ ring a 6? electron aromatic system. Also, note that, for M=K, the BLA is maximum in the series consistent with the smallest net aromaticity as supported by NICS and HOMO-LUMO gap calculations. The stability of the complexes is measured using the fragmentation scheme: $(Al_4M_4)$-$Fe(CO)_3$ = $Al_4M_4$ + $Fe(CO)_3$; where the constituents are in their stable ground state geometric configurations. The binding energies [defined as $E_{complex}$-$E_{fragments}$] in $(Al_4Li_4)$-$Fe(CO)_3$, $(Al_4Na_4)$-$Fe(CO)_3$ and $(Al_4K_4)$-$Fe(CO)_3$ are -118.85 kcal/mol, -122.92 kcal/mol and -126.28 kcal/mol respectively. For comparison, we have also calculated the binding energies in $(C_4H_4)$-$Fe(CO)_3$, which is -70 kcal/mol.

The large stability associated with these molecules is due to the formation of a stable closed shell 18-e configuration in each case.[14]

The binding energy for $(Al_4K_4)$-$Fe(CO)_3$ is maximum suggesting that in such complexes, only the ?-electrons on the ring interact with the metal 3d orbitals and the poor ?-? separation initially existing in the ligand is lifted and the system behaves as an 18 electron stable molecule. This is a very important result as one can fine tune the ?-? separations by functionalizing the $Al_4^{4-}$ rings with electropositive counterions. The magnitudes for the HOMO-LUMO gaps for the $(Al_4Li_4)$-$Fe(CO)_3$, $(Al_4Na_4)$-$Fe(CO)_3$ and $(Al_4K_4)$-$Fe(CO)_3$ complexes decrease in the order : 3.34 eV, 2.31 eV and 1.85 eV respectively, suggesting the softer nature of the $Al_4K_4$ ligand. Another major change that is associated with the formation of the complexes is that the structure of the counterions are lost completely. As already mentioned, the counterions are arranged around the $Al_4$ ring so as to have a $C_{2h}$ symmetry. However, on complexation with the transition metal atoms, the symmetry is completely lost. There is an overall loss in energy of the order 20 kcal/mol, compared to the ground state structure of $Al_4M_4$. But, the stability associated with the formation of the complex with the 3d metal atom overwhelms the structural instability in the ligand.

Another very well-known route for stabilizing the unstable ligands is through the formation of sandwich complexes similar to that in the case of ferrocene.[15] For 4? electronic systems like $C_4H_4$, $(C_4H_4)_2Ni$ is a well stabilized complex.[16] Similarly, we have stabilized sandwich complexes of the type: $(Al_4M_4)_2Ni$ having binding energies similar to that for the organic analogues. The binding energies for $(Al_4Li_4)_2Ni$, $(Al_4Na_4)_2Ni$ and $(Al_4K_4)_2Ni$ are -146.05 kcal/mol, -147.12 kcal/mol and -103.12 kcal/mol respectively. Note that, lower stabilization in $(Al_4K_4)_2Ni$ arises from the distortion in the sandwich architecture due to the presence of the bulky $K^+$ ions (see structure in Scheme 1) as a result of which the average $K^+$ ion distance to the $Al_4^{4-}$ ring is very large (3.5 Å). For $(Al_4Li_4)_2Ni$ and $(Al_4Na_4)_2Ni$, the average $M^+$ distance from the $Al_4^{4-}$ ring is 3.0 Å. This is also evident in the decreasing order of the HOMO-LUMO gaps ( 1.623 eV, 1.323 eV and 0.954 eV for $(Al_4Li_4)_2Ni$, $(Al_4Na_4)_2Ni$ and $(Al_4K_4)_2Ni$ respectively). The binding energy for $(C_4H_4)_2Ni$ is -160.42 kcal/mol and thus unlike the cases for $(Al_4M_4)$-$Fe(CO)_3$, direct substitution of $C_4H_4$ with $Al_4M_4$ will be highly endothermic and thus quite unfavourable. For a detailed understanding of the highly exothermic formation of $(Al_4M_4)$-$Fe(CO)_3$ compared to the endothermic substitution product $(Al_4M_4)_2Ni$, the HOMO orbitals for both the systems have been analysed (see supporting information file). The HOMO for $(Al_4M_4)$-$Fe(CO)_3$ shows substantially more intermixing between the d-orbitals of the $Fe(CO)_3$ fragment and the ?- orbitals of Al$_4$M$_4$, leading to stronger complexation in the case of (Al$_4$M$_4$)-Fe(CO)$_3$. The presence of three strong π-acceptor CO ligands in the Fe(CO)$_3$ fragment leads to quenching of d-orbitals on the Fe-atom and thereby facilitates stronger binding between the Al$_4$M$_4$ ligand and the Fe(CO)$_3$ fragment.

In view of the above, we follow here a different strategy. We consider a substitution reaction of the type: (C4H4)2Ni + Al4M4 = (C4H4)Ni (Al4M4) + C4H4. As already mentioned, the Al4M4 binds stronger to the metal center than C4H4. Therefore, we expect that it is possible to synthesise a hybrid organic-inorganic sandwich complex. These hybrid complexes are very interesting because, while for the C4H4 ligand, the interaction with the transition metal atom involves only the π electrons, for the Al4M4 ligand, the interaction are through both the σ and π orbitals. Interestingly, the extent of the involvement of the σ and π orbitals in the interaction does also depend on the nature of the counter ions present. In Fig. 3, we show the frontier orbital plots for (C4H4)Ni(Al4Li4). Similar features are also seen in the cases of (C4H4)Ni(Al4Na4) and (C4H4)Ni(Al4K4). From these orbitals, it is clear that the ligand group orbitals (LGOs) on Al4M4 interact much strongly with the transition metal (Ni) orbitals. Specifically, the dZ2, dxz and dxy orbitals of Ni interact with the LGOs of Al4M4 in HOMO-1, H0M0-2 and HOMO-3 respectively. On the contrary, the LGOs of C4H4 interact very feebly with the transition metal orbitals, as a result of which, the binding of the organic ligand with the transition metal is much weaker compared to that for Al4M4 in these mixed sandwich complexes. The structures of these hybrid complexes are found to be quite stable. The heat of formation for (C4H4)Ni (Al4Li4), (C4H4)Ni (Al4Na4) and (C4H4)Ni (Al4K4) are -153.93 kcal/mol, -158.82 kcal/mol and -151.80 respectively. This is in marked contrast to that for the previous complexes where the heat of formation for the complexes increased monotonically with the increase in the sizes of the counterions (Fig. 3). The HOMO-LUMO gaps in the hybrid complexes, however, follow the same trend as other complexes: 2.01 eV [(C4H4)Ni (Al4Li4)], 1.96 eV [(C4H4)Ni (Al4Na4)] and 1.35 eV [(C4H4)Ni (Al4K4)].

**Figure 3** Frontier orbital diagrams in mixed sandwich complex, (Al$_4$Li$_4$)Ni(C$_4$H$_4$).

This explains smaller binding energy in the hybrid complex with K+ as the counterion compared to (C4H4)Ni (Al4Li4) and(C4H4)Ni(Al4Na4). However, compared to the Fe(CO)3 complexes, these complexes are softer as the HOMO-LUMO gap is quite small.

We would like to rationalize the synthesis of these sandwich complexes in a 3 step reaction of the type: (C$_4$H$_4$)Ni(C$_4$H$_4$) to (C$_4$H$_4$)Ni (Al$_4$M$_4$) and finally to (Al$_4$M$_4$)Ni (Al$_4$M$_4$) [shown in Scheme. 1].

As already shown that the intermediates, (C$_4$H$_4$)Ni (Al$_4$M$_4$), are quite stable and can thus be isolated. However, these substitution reactions are mildly endothermic (contrary to that for the Fe(CO)3 complexes) In this series, the heat of formation is least endothermic for both (Al$_4$Na$_4$)Ni(C$_4$H$_4$) and (Al$_4$Na$_4$)$_2$Ni and thus are best candidates for isolation.

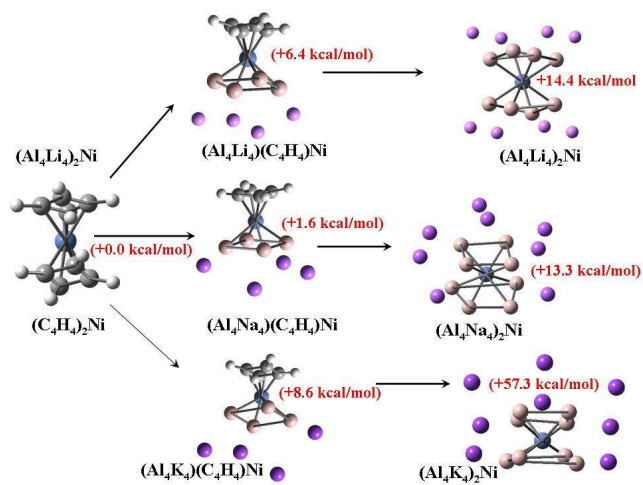

**Scheme1:** Stepwise synthesis for all-metal sandwich complexes from organometallic complex (C$_4$H$_4$)$_2$Ni. The energy for (C$_4$H$_4$)$_2$Ni has been scaled to zero to show the endothermic substitution reactions. Small filled circles indicate the alkali metal ions.

The BLA for C4H4 and Al4M4 are 0.0092 and 0.0386 in (C4H4)Ni (Al4Li4), 0.0106 and 0.0168 in (C4H4)Ni (Al4Na4) and 0.0033 and 0.10459 in (C4H4)Ni (Al4K4) respectively. Note that, for M=K, the Al4K4 unit has a substantial BLA with a magnitude close to that of uncoordinated Al4K4. To conclude, we propose experimental methodologies for possible synthesis of the first examples of all-metal half-sandwich, mixed sandwich and full-sandwich complexes. Based on our high level DFT calculations, we have shown that while half-sandwich complexes can be readily formed due to the presence of π-accepting ligands like CO, the formation of mixed and full-sandwich complexes are endothermic.

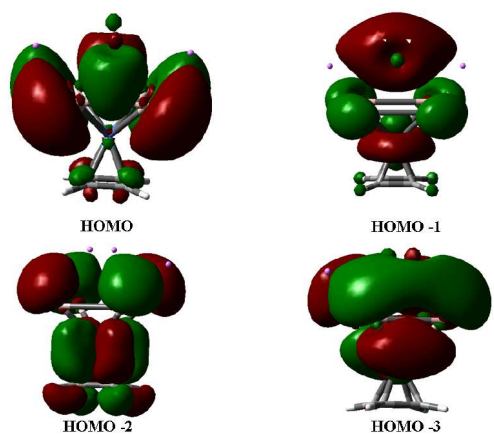

SKP thanks DST and CSIR govt. of India for financial support.

**Ayan Datta and Swapan K. Pati**[a]
[a] *Chemistry and Physics of Materials Unit and Theoretical Sciences Unit Jawaharlal Nehru Center for Advcanced Scientific Research, Jakkur P.O, Bangalore, India. Fax: +91-080-846766 ; Tel: +91-080-8462750 ; E-mail: pati@jncasr.ac.in*

## Notes and references